# SIGNATURES OF UNIVERSAL CHARACTERISTICS OF FRACTAL FLUCTUATIONS IN GLOBAL MEAN MONTHLY TEMPERATURE ANOMALIES

A. M. SELVAM[1]

*Running head*: Universal characteristics of fractal fluctuations

**Abstract** Selfsimilar space-time fractal fluctuations are generic to dynamical systems in nature such as atmospheric flows, heartbeat patterns, population dynamics, etc. The physics of the long-range correlations intrinsic to fractal fluctuations is not completely understood. It is important to quantify the physics underlying the irregular fractal fluctuations for prediction of space-time evolution of dynamical systems. A general systems theory for fractals visualising the emergence of successively larger scale fluctuations resulting from the space-time integration of enclosed smaller scale fluctuations is proposed. The theoretical model predictions are: (i) The probability distribution and the power spectrum for fractal fluctuations is the same inverse power law function incorporating the *golden mean*. (ii) The predicted distribution is close to the Gaussian distribution for small-scale fluctuations but exhibits *fat long tail* for large-scale fluctuations with higher probability of occurrence than predicted by Gaussian distribution. (iii) Since the power spectrum (variance, i.e., square of eddy amplitude) also represents the probability densities as in the case of quantum systems such as the electron or photon, fractal fluctuations exhibit quantumlike chaos. (iv) The *fine structure constant* for spectrum of fractal fluctuations is a function of the golden mean and is analogous to atomic spectra equal to about 1/137. Global gridded time series data sets of monthly mean temperatures for the period 1880 – 2007/2008 were analysed. The data sets and the corresponding power spectra exhibit distributions close to the model predicted inverse power law distribution. The model predicted and observed universal spectrum for interannual variability rules out linear secular trends in global monthly mean temperatures. Global warming results in intensification of fluctuations of all scales and manifested immediately in high frequency fluctuations.

## 1 Introduction

Dynamical systems in nature such as atmospheric flows, heartbeat patterns, population dynamics, stock market indices, DNA base A, C, G, T sequence pattern, etc., exhibit irregular space-time fluctuations on all scales and exact quantification of the fluctuation pattern for predictability purposes has not yet been achieved. Traditionally, the Gaussian probability distribution is used for a broad quantification of the data set variability in terms of the sample mean and variance. The fractal or selfsimilar nature of space-time fluctuations was identified by Mandelbrot[1] in the 1970s. Fractal fluctuations show a zigzag selfsimilar pattern of successive increase

[1] Deputy Director (Retired), Indian Institute of Tropical Meteorology, Pune 411 008, India, Email: amselvam@gmail.com, Web sites: http://amselvam.webs.com, http://amselvam.tripod.com/index.html, Address for communication: Dr.A.M.Selvam, B1 Aradhana, 42/2A Shivajinagar, Pune 411 005, India, Res. Ph: 091-020-25538194

followed by decrease on all scales (space-time), for example in atmospheric flows, cycles of increase and decrease in meteorological parameters such as wind, temperature, etc. occur from the turbulence scale of millimeters-seconds to climate scales of thousands of kilometers-years. The power spectra of fractal fluctuations exhibit inverse power law of the form $f^{\alpha}$ where $f$ is the frequency and $\alpha$ is a constant. Inverse power law for power spectra indicate long-range space-time correlations or scale invariance for the scale range for which $\alpha$ is a constant, i.e., the amplitudes of the eddy fluctuations in this scale range are a function of the scale factor $\alpha$ alone. In general the value of $\alpha$ is different for different scale ranges indicating multifractal structure for the fluctuations. The long-range space-time correlations exhibited by dynamical systems are identified as self-organized criticality[2-3]. The physics of self-organized criticality is not yet identified. The physics of fractal fluctuations generic to dynamical systems in nature is not yet identified and traditional statistical, mathematical theories do not provide adequate tools for identification and quantitative description of the observed universal properties of fractal structures observed in all fields of science and other areas of human interest. A recently developed general systems theory for fractal space-time fluctuations[4-7] shows that the larger scale fluctuation can be visualized to emerge from the space-time averaging of enclosed small scale fluctuations, thereby generating a hierarchy of selfsimilar fluctuations manifested as the observed eddy continuum in power spectral analyses of fractal fluctuations. Such a concept results in inverse power law form incorporating the golden mean $\tau$ for the space-time fluctuation pattern and also for the power spectra of the fluctuations (Section 3). The predicted distribution is close to the Gaussian distribution for small-scale fluctuations, but exhibits *fat long tail* for large-scale fluctuations. Analyses of extensive data sets of Global gridded data sets of monthly mean temperatures for the period 1880 – 2007/2008 show that the space/time data sets follow closely, but not exactly the statistical normal distribution, particularly in the region of normalized deviations $t$ greater than 2, the $t$ values being computed as equal to $(x\text{-}av)/sd$ where $av$ and $sd$ denote respectively the mean and standard deviation of the variable $x$. The general systems theory, originally developed for turbulent fluid flows, provides universal quantification of physics underlying fractal fluctuations and is applicable to all dynamical systems in nature independent of its physical, chemical, electrical, or any other intrinsic characteristic. In the following, Section 2 gives a summary of traditional statistical and mathematical theories/techniques used for analysis and quantification of space-time fluctuation data sets. The general systems theory for fractal space-time fluctuations is described in Section 3. The Boltzmann distribution of classical statistical physics is discussed in the context of general systems theory concepts in Section 4. Section 5 deals with data and analyses techniques. Discussion and conclusions of results are presented in Section 6.

## 2 Statistical Methods for Data Analysis

Dynamical systems such as atmospheric flows, stock markets, heartbeat patterns, population growth, traffic flows, etc., exhibit irregular space-time fluctuation patterns. Quantification of the space-time fluctuation pattern will help predictability studies, in particular for events which affect day-to-day human life such as extreme weather events, stock market crashes, traffic jams, etc. The analysis of data sets and broad quantification in terms of probabilities belongs to the field of statistics. Early attempts resulted in identification of two quantitative (mathematical) distributions which approximately fit data sets from a wide range of scientific and other disciplines of study. The first is the well known statistical normal distribution and the second is the power law distribution associated with the recently identified ‘*fractals*’ or selfsimilar characteristic of data sets in general.

### 2.1 Statistical Normal Distribution

Most quantitative research involves the use of statistical methods presuming *independence* among data points and Gaussian ‘normal’ distributions[8]. The Gaussian distribution is reliably characterized by its stable mean and finite variance[9]. Even the largest deviations, which are exceptionally rare, are still only about a factor of two from the mean in either direction and are well characterized by quoting a simple standard deviation[10]. However, apparently rare real life catastrophic events such as major earth quakes, stock market crashes, heavy rainfall events, etc., occur more frequently than indicated by the normal curve, i.e., they exhibit a probability distribution with a *fat tail*. Fat tails indicate a power law pattern and interdependence. The “tails” of a power-law curve — the regions to either side that correspond to large fluctuations — fall off very slowly in comparison with those of the bell curve[11]. The normal distribution is therefore an inadequate model for extreme departures from the mean.

### 2.2 Fractal Fluctuations and Statistical Analysis

Fractals are the latest development in statistics. The space-time fluctuation pattern in dynamical systems was shown to have a selfsimilar or fractal structure in the 1970s[1]. Representative examples of fractal fluctuations of monthly mean temperatures for the period 1880 to 2007/2008 are shown Figure 1.

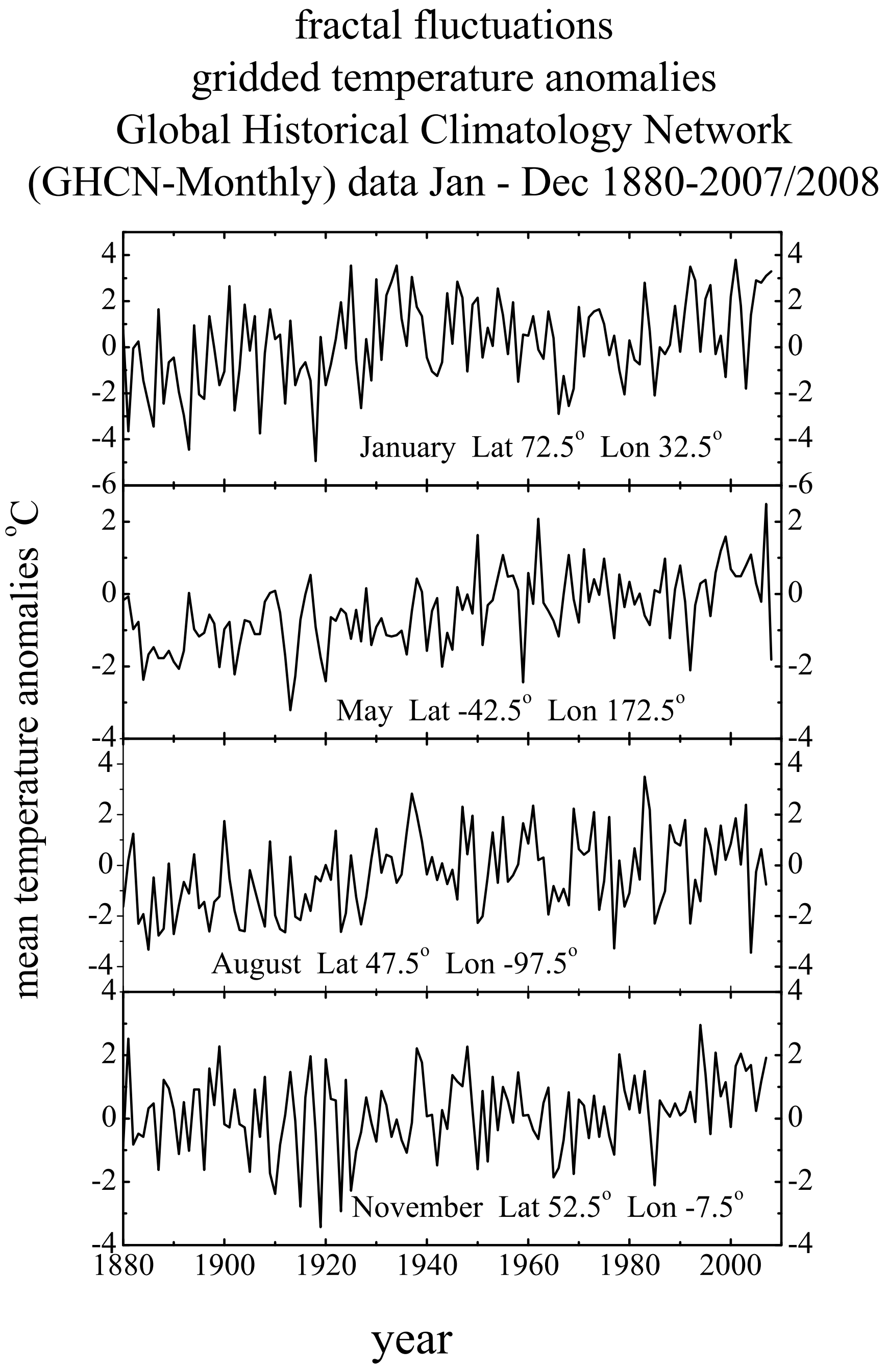

Figure 1 Representative examples of fractal fluctuations of Global gridded monthly mean temperature anomalies for the period 1880 – 2007/2008

The larger scale fluctuation consists of smaller scale fluctuations identical in shape to the larger scale. An appreciation of the properties of fractals is changing the most basic ways we analyze and interpret data from experiments and is leading to new insights into understanding physical, chemical, biological, psychological, and social systems. The selfsimilar fluctuations imply long-range space-time correlations or interdependence. Therefore, the Gaussian distribution will not be applicable for description of fractal data sets. However, the bell curve still continues to be used for approximate quantitative characterization of data which are now identified as fractal space-time fluctuations.

### 2.2.1 Power laws and fat tails

Fractals conform to power laws. A power law is a relationship in which one quantity $A$ is proportional to another $B$ taken to some power $n$; that is, $A \sim B^n$[11]. Andriani and McKelvey[8] have given exhaustive references to earliest known work on power law relationships. Selvam[12] has summarized earlier reported power law relationships in dynamical systems.

A power law distribution indicates the absence of a characteristic size and as a consequence there is no upper limit on the size of events[13]. A power law world is dominated by extreme events ignored in a Gaussian-world. In fact, the fat tails of power law distributions make large extreme events orders-of-magnitude more likely. Theories explaining power laws are also scale-free. This is to say, the same explanation (theory) applies at all levels of analysis[8].

### 2.2.2 Scale-free theory for power laws with fat, long tails

A scale-free theory for the observed fractal fluctuations in atmospheric flows shows that the observed long-range spatiotemporal correlations are intrinsic to quantumlike chaos governing fluid flows. The model concepts are independent of the exact details such as the chemical, physical, physiological and other properties of the dynamical system and therefore provide a general systems theory applicable to all real world and computed dynamical systems in nature[4-7,14-20]. The model is based on the concept that the irregular fractal fluctuations may be visualized to result from the superimposition of an eddy continuum, i.e., a hierarchy of eddy circulations generated at each level by the space-time integration of enclosed small-scale eddy fluctuations. Such a concept of space-time fluctuation averaged distributions *should* follow statistical normal distribution according to *Central Limit Theorem* in traditional Statistical theory[21]. Also, traditional statistical/mathematical theory predicts that the Gaussian, its Fourier transform and therefore Fourier transform associated power spectrum are the same distributions. The Fourier transform of normal distribution is essentially a normal distribution. A power spectrum is based on the Fourier transform, which expresses the relationship between time (space) domain and frequency domain description of any physical process[22-23]. However, the model (Section 3) visualises the eddy growth process in successive stages of unit length-step growth with ordered two-way energy feedback between the larger and smaller scale eddies and derives a power law probability distribution $P$ which is close to the Gaussian for small deviations and gives the observed fat, long tail for large fluctuations. Further, the model predicts the power spectrum of the eddy continuum also to follow the power law probability distribution $P$.

In summary, the model predicts the following:

- The eddy continuum consists of an overall logarithmic spiral trajectory with the quasiperiodic *Penrose* tiling pattern for the internal structure.
- The successively larger eddy space-time scales follow the Fibonacci number series.
- The probability distribution $P$ of fractal domains for the $n^{th}$ step of eddy growth is equal to $\tau^{-4n}$ where $\tau$ is the golden mean equal to $(1+\sqrt{5})/2$ ($\approx 1.618$).

The eddy growth step *n* represents the *normalized deviation t* in traditional statistical theory. The *normalized deviation t* represents the departure of the variable from the mean in terms of the standard deviation of the distribution assumed to follow *normal distribution* characteristics for many real world space-time events. There is progressive decrease in the probability of occurrence of events with increase in corresponding *normalized deviation t*. Space-time events with *normalized deviation t* greater than 2 occur with a probability of 5 percent or less and may be categorized as extreme events associated in general with widespread (space-time) damage and loss. The model predicted probability distribution *P* is close to the statistical normal distribution for *t* values less than 2 and greater than normal distribution for *t* more than 2, thereby giving a *fat, long tail*. There is non-zero probability of occurrence of very large events.

- The inverse of probability distribution *P*, namely, $\tau^{4n}$ represents the relative eddy energy flux in the large eddy fractal (small scale fine structure) domain. There is progressive decrease in the probability of occurrence of successive stages of eddy growth associated with progressively larger domains of fractal (small scale fine structure) eddy energy flux and at sufficiently large growth stage trigger catastrophic extreme events such as heavy rainfall, stock market crashes, traffic jams, etc., in real world situations.
- The power spectrum of fractal fluctuations also follows the same distribution *P* as for the distribution of fractal fluctuations. The square of the eddy amplitude (variance) represents the eddy energy and therefore the eddy probability density *P*. Such a result that the additive amplitudes of eddies when squared represent probabilities, is exhibited by the sub-atomic dynamics of quantum systems such as the electron or proton[24-26]. Therefore fractal fluctuations are signatures of quantumlike chaos in dynamical systems.
- The *fine structure constant* for spectrum of fractal fluctuations is a function of the *golden mean* and is analogous to that of atomic spectra equal to about 1/137.
- The universal algorithm for self-organized criticality is expressed in terms of the universal *Feigenbaum*'s constants[27] *a* and *d* as $2a^2 = \pi d$ where the fractional volume intermittency of occurrence $\pi d$ contributes to the total variance $2a^2$ of fractal structures. The *Feigenbaum*'s constants are expressed as functions of the *golden mean*. The probability distribution *P* of fractal domains is also expressed in terms of the Feigenbaum's constants *a* and *d*. The details of the model are summarized in the following section (Section 3).

## 3 A General Systems Theory for Fractal Fluctuations

The fractal space-time fluctuations of dynamical systems may be visualized to result from the superimposition of an ensemble of eddies (sine waves), namely an eddy continuum. The relationship between large and small eddy circulation parameters are obtained on the basis of Townsend's[28] concept that large eddies are envelopes enclosing turbulent eddy (small-scale) fluctuations (Figure 2).

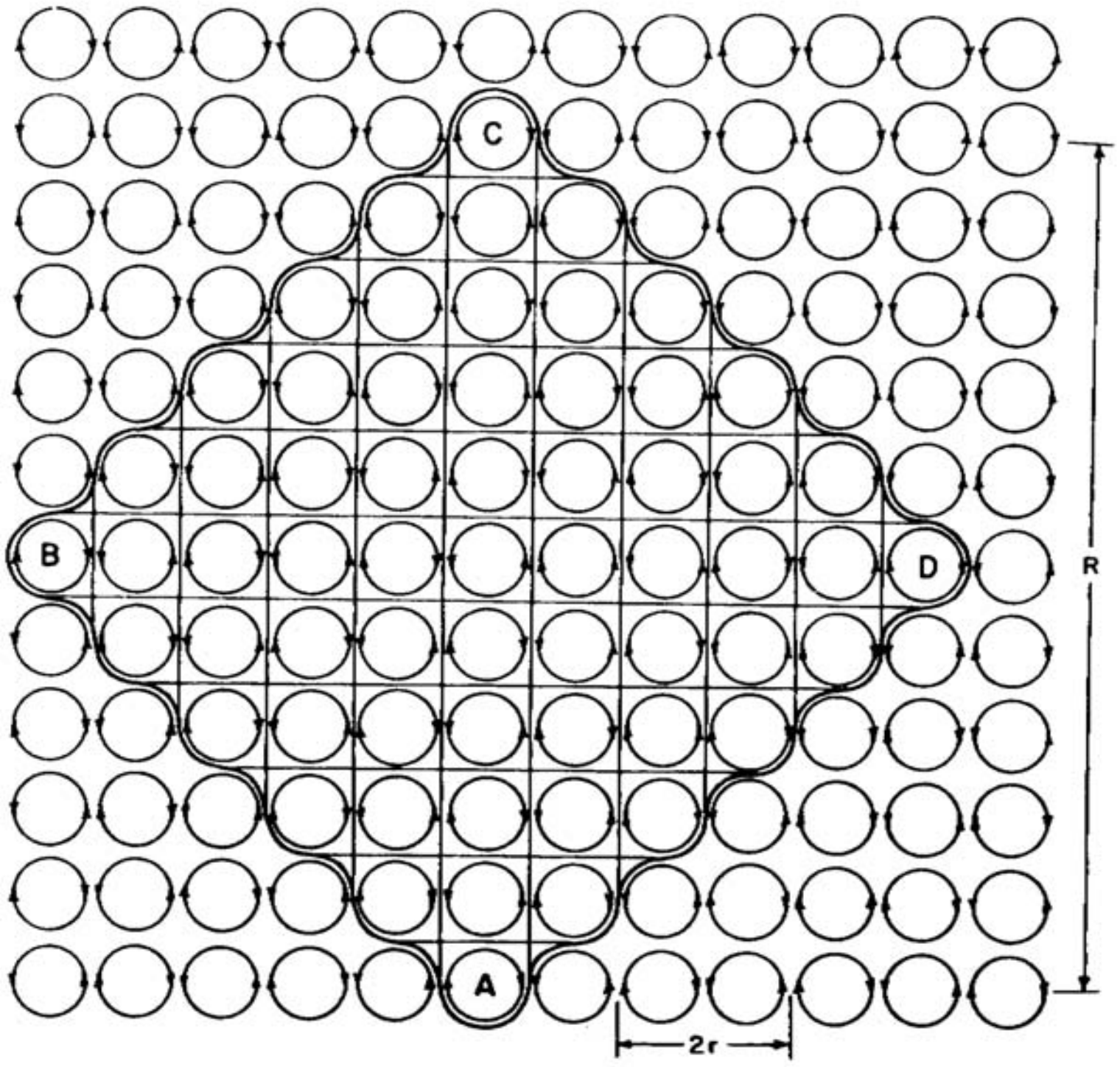

Figure 2 Physical concept of eddy growth process by the self-sustaining process of ordered energy feedback between the larger and smaller scales, the smaller scales forming the internal circulations of larger scales. The figure shows a uniform distribution of dominant turbulent scale eddies of length scale 2r. Larger-eddy circulations such as ABCD form as coherent structures sustained by the enclosed turbulent eddies.

The relationship between root mean square (r. m. s.) circulation speeds $W$ and $w_*$ respectively of large and turbulent eddies of respective radii $R$ and $r$ is then given as

$$W^2 = \frac{2}{\pi}\frac{r}{R} w_*^2 \tag{1}$$

The dynamical evolution of space-time fractal structures is quantified in terms of ordered energy flow between fluctuations of all scales in equation (1), because the square of the eddy circulation speed represents the eddy energy (kinetic). A hierarchical continuum of eddies is generated by the integration of successively larger enclosed turbulent eddy circulations. Such a concept of space-time fluctuation averaged distributions *should* follow statistical normal distribution according to *Central Limit Theorem* in traditional Statistical theory[21]. Also, traditional statistical/mathematical theory predicts that the Gaussian, its Fourier transform and therefore Fourier transform associated power spectrum are the same distributions. However, the general systems theory[4-7,14-20] visualises the eddy growth process in successive stages of unit length-step growth with ordered two-way energy feedback between the larger and smaller scale eddies and derives a power law probability distribution $P$ which is close to the Gaussian for small deviations and gives the observed fat, long tail for large fluctuations. Further, the model predicts the power spectrum of the eddy continuum also to follow the power law probability distribution $P$. Therefore the additive amplitudes of the eddies when squared (variance), represent the probability distribution similar to the subatomic dynamics of quantum systems such as the electron or photon. Fractal fluctuations therefore exhibit quantumlike chaos.

The above-described analogy of quantumlike mechanics for dynamical systems is similar to the concept of a subquantum level of fluctuations whose space-time organization gives rise to the observed manifestation of subatomic phenomena, i.e., quantum systems as order out of chaos phenomena[29].

### 3.1 Quasicrystalline Structure of the Eddy Continuum

The turbulent eddy circulation speed and radius increase with the progressive growth of the large eddy[4-6]. The successively larger turbulent fluctuations, which form the internal structure of the growing large eddy, may be computed (equation 1) as

$$w_*^2 = \frac{\pi}{2}\frac{R}{\mathrm{d}R}W^2 \qquad (2)$$

During each length step growth d$R$, the small-scale energizing perturbation $W_n$ at the $n^{\text{th}}$ instant generates the large-scale perturbation $W_{n+1}$ of radius $R$ where $R = \sum_1^n \mathrm{d}R$ since successive length-scale doubling gives rise to $R$. Equation 2 may be written in terms of the successive turbulent circulation speeds $W_n$ and $W_{n+1}$ as

$$W_{n+1}^2 = \frac{\pi}{2}\frac{R}{\mathrm{d}R}W_n^2 \qquad (3)$$

The angular turning d$\theta$ inherent to eddy circulation for each length step growth is equal to d$R$/$R$. The perturbation d$R$ is generated by the small-scale acceleration $W_n$ at any instant $n$ and therefore d$R$=$W_n$. Starting with the unit value for d$R$ the successive $W_n$, $W_{n+1}$, $R$, and d$\theta$ values are computed from equation 3 and are given in Table 1.

Table 1 The computed spatial growth of the strange-attractor design traced by the macro-scale dynamical system of atmospheric flows as shown in Figure 3.

| $R$ | $W_n$ | d$R$ | d$\theta$ | $Wn$+1 | $\theta$ |
|---|---|---|---|---|---|
| 1.000 | 1.000 | 1.000 | 1.000 | 1.254 | 1.000 |
| 2.000 | 1.254 | 1.254 | 0.627 | 1.985 | 1.627 |
| 3.254 | 1.985 | 1.985 | 0.610 | 3.186 | 2.237 |
| 5.239 | 3.186 | 3.186 | 0.608 | 5.121 | 2.845 |
| 8.425 | 5.121 | 5.121 | 0.608 | 8.234 | 3.453 |
| 13.546 | 8.234 | 8.234 | 0.608 | 13.239 | 4.061 |
| 21.780 | 13.239 | 13.239 | 0.608 | 21.286 | 4.669 |
| 35.019 | 21.286 | 21.286 | 0.608 | 34.225 | 5.277 |
| 56.305 | 34.225 | 34.225 | 0.608 | 55.029 | 5.885 |
| 90.530 | 55.029 | 55.029 | 0.608 | 88.479 | 6.493 |

It is seen that the successive values of the circulation speed $W$ and radius $R$ of the growing turbulent eddy follow the Fibonacci mathematical number series such that $R_{n+1}=R_n+R_{n-1}$ and $R_{n+1}/R_n$ is equal to the golden mean $\tau$, which is equal to $[(1 + \sqrt{5})/2] \cong (1.618)$. Further, the successive $W$ and $R$ values form the geometrical progression $R_0(1+\tau+\tau^2+\tau^3+\tau^4+ \ldots)$ where $R_0$ is the initial value of the turbulent eddy radius.

Turbulent eddy growth from primary perturbation $OR_O$ starting from the origin O (Figure 3) gives rise to compensating return circulations $OR_1R_2$ on either side of $OR_O$, thereby generating the large eddy radius $OR_1$ such that $OR_1/OR_O=\tau$ and $R_OOR_1=\pi/5=R_OR_1O$. Therefore, short-range circulation balance requirements generate successively larger circulation patterns with precise geometry that is governed by the *Fibonacci* mathematical number series, which is identified as a signature of the universal period doubling route to chaos in fluid flows, in particular atmospheric flows. It is seen from Figure 3 that five such successive length step growths give successively increasing radii $OR_1$, $OR_2$, $OR_3$, $OR_4$ and $OR_5$ tracing out one complete vortex-roll circulation such that the scale ratio $OR_5/OR_O$ is equal to $\tau^5=11.1$. The envelope $R_1R_2R_3R_4R_5$ (Figure 3) of a dominant large eddy (or vortex roll) is found to fit the logarithmic spiral $R=R_0e^{b\theta}$ where $R_0=OR_O$, $b=\tan\delta$ with $\delta$ the crossing angle equal to $\pi/5$, and the angular turning $\theta$ for each length step growth is equal to $\pi/5$. The successively larger eddy radii may be subdivided again in the *golden mean* ratio. The internal structure of large-eddy circulations is therefore made up of balanced small-scale circulations tracing out the well-known quasi-periodic *Penrose* tiling pattern identified as the quasi-crystalline structure in condensed matter physics. A complete description of the atmospheric flow field is given by the quasi-periodic cycles with *Fibonacci* winding numbers.

### 3.2 Model Predictions

The model predictions[4-7] are

#### 3.2.1 Quasiperiodic Penrose tiling pattern

Atmospheric flows trace an overall logarithmic spiral trajectory $OR_OR_1R_2R_3R_4R_5$ simultaneously in clockwise and anti-clockwise directions with the quasi-periodic *Penrose tiling pattern*[30] for the internal structure shown in Figure 3.

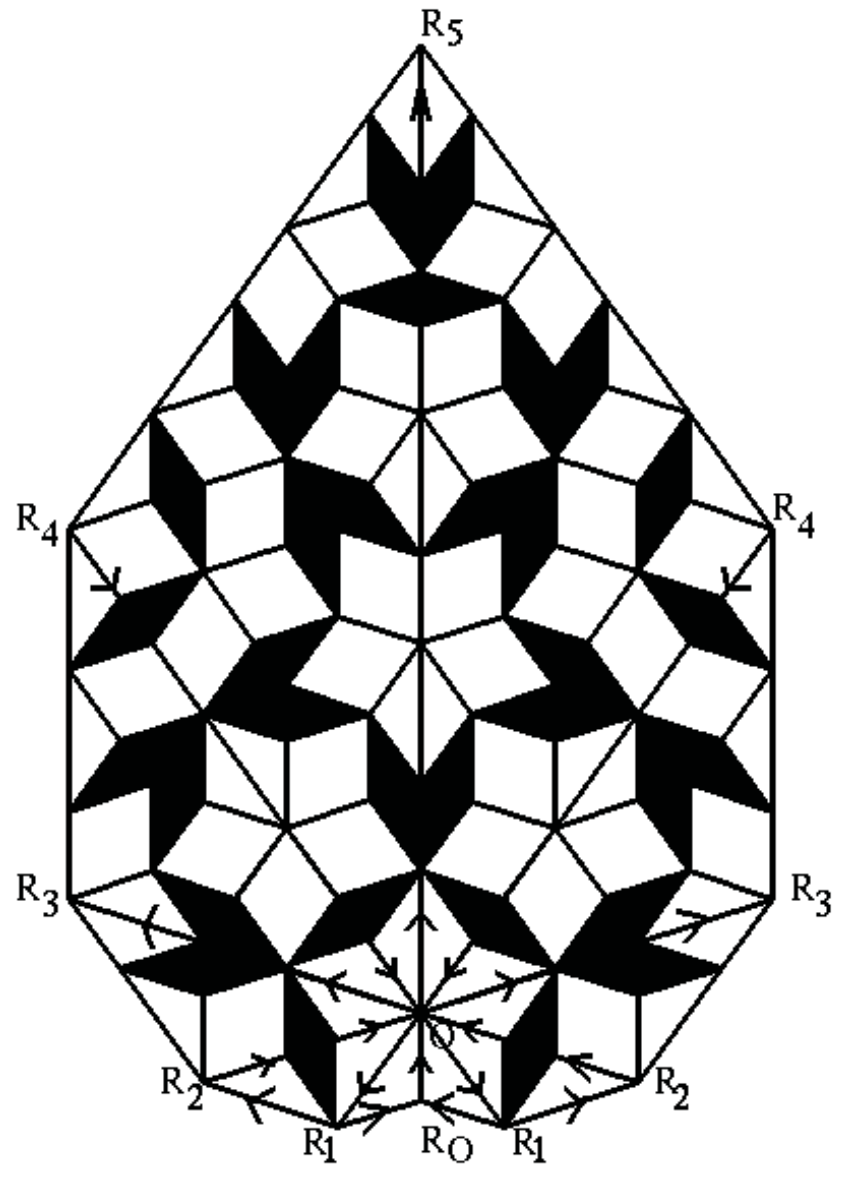

Figure 3 The quasiperiodic *Penrose* tiling pattern

The spiral flow structure can be visualized as an eddy continuum generated by successive length step growths $OR_O$, $OR_1$, $OR_2$, $OR_3$,….respectively equal to $R_1$, $R_2$, $R_3$,….which follow *Fibonacci* mathematical series such that $R_{n+1}=R_n+R_{n-1}$ and $R_{n+1}/R_n=\tau$ where $\tau$ is the *golden mean* equal to $(1+\sqrt{5})/2$ ($\approx 1.618$). Considering a normalized length step equal to 1 for the last stage of eddy growth, the successively decreasing radial length steps can be expressed as 1, $1/\tau$, $1/\tau^2$, $1/\tau^3$, ……The normalized eddy continuum comprises of fluctuation length scales 1, $1/\tau$, $1/\tau^2$, …….. The probability of occurrence is equal to $1/\tau$ and $1/\tau^2$ respectively for eddy length scale $1/\tau$ in any one or both rotational (clockwise and anti-clockwise) directions. Eddy fluctuation length of amplitude $1/\tau$ has a probability of occurrence equal to $1/\tau^2$ in both rotational directions, i.e., the square of eddy amplitude represents the probability of occurrence in the eddy continuum. Similar result is observed in the subatomic dynamics of quantum systems which are visualized to consist of the superimposition of eddy fluctuations in wave trains (eddy continuum).

### 3.2.2 Logarithmic spiral pattern underlying fractal fluctuations

The overall logarithmic spiral flow structure (Figure 3) is given by the relation

$$W = \frac{w_*}{k} \ln z \qquad (4)$$

In equation 4 the constant $k$ is the steady state fractional volume dilution of large eddy by inherent turbulent eddy fluctuations and $z$ is the length scale ratio $R/r$. The constant $k$ is equal to $1/\tau^2$ ($\cong 0.382$) and is identified as the universal constant for deterministic chaos in fluid flows. The steady state emergence of fractal structures is therefore equal to

$$\frac{1}{k} \cong 2.62 \qquad (5)$$

In equation 4, $W$ represents the standard deviation of eddy fluctuations, since $W$ is computed as the instantaneous r. m. s. (root mean square) eddy perturbation amplitude with reference to the earlier step of eddy growth. For two successive stages of eddy growth starting from primary perturbation $w_*$, the ratio of the standard deviations $W_{n+1}$ and $W_n$ is given from equation 4 as $(n+1)/n$. Denoting by $\sigma$ the standard deviation of eddy fluctuations at the reference level ($n=1$) the standard deviations of eddy fluctuations for successive stages of eddy growth are given as integer multiples of $\sigma$, i.e., $\sigma$, $2\sigma$, $3\sigma$, etc. and correspond respectively to

$$\text{statistical normalised standard deviation} \;\; t = 0, 1, 2, 3, .... \qquad (6)$$

The conventional power spectrum plotted as the variance versus the frequency in log-log scale will now represent the eddy probability density on logarithmic scale versus the standard deviation of the eddy fluctuations on linear scale since the logarithm of the eddy wavelength represents the standard deviation, i.e., the r. m. s. value of eddy fluctuations (equation 4). The r. m. s. value of eddy fluctuations can be represented in terms of statistical normal distribution as follows. A normalized standard deviation $t=0$ corresponds to cumulative percentage probability density equal to 50 for the mean value of the distribution. Since the logarithm of the wavelength represents the r. m. s. value of eddy fluctuations the normalized standard deviation $t$ is defined for the eddy energy as

$$t = \frac{\log L}{\log T_{50}} - 1 \qquad (7)$$

In equation 7 $L$ is the time period (or wavelength) and $T_{50}$ is the period up to which the cumulative percentage contribution to total variance is equal to 50 and $t = 0$. Log$T_{50}$ also represents the mean value for the r. m. s. eddy fluctuations and is consistent with the concept of the mean level represented by r. m. s. eddy fluctuations. Spectra of time series of meteorological parameters when plotted as cumulative percentage contribution to total variance versus $t$ have been shown to follow closely the model predicted universal spectrum[7] which is identified as a signature of quantumlike chaos.

### 3.2.3 Universal Feigenbaum's constants and probability distribution function for fractal fluctuations

Selvam[31,6] has shown that equation 1 represents the universal algorithm for deterministic chaos in dynamical systems and is expressed in terms of the universal *Feigenbaum*'s *Feigenbaum*'s[27] *constants a* and *d* as follows. The successive length step growths generating the eddy continuum $OR_OR_1R_2R_3R_4R_5$ (Figure 3) analogous to the period doubling route to chaos (growth) is initiated and sustained by the turbulent (fine scale) eddy acceleration $w_*$, which then propagates by the inherent property of inertia of the medium of propagation. Therefore, the statistical parameters *mean*, *variance*, *skewness* and *kurtosis* of the perturbation field in the medium of propagation are given by $w_*$, $w_*^2$, $w_*^3$ and $w_*^4$ respectively. The associated dynamics of the perturbation field can be described by the following parameters. The perturbation speed $w_*$ (motion) per second (unit time) sustained by its inertia represents the mass, $w_*^2$ the acceleration or force, $w_*^3$ the angular momentum or potential energy, and $w_*^4$ the spin angular momentum, since an eddy motion has an inherent curvature to its trajectory.
It is shown that *Feigenbaum's* constant $a$ is equal to[31, 6]

$$a = \frac{W_2 R_2}{W_1 R_1} \tag{8}$$

In equation 8 the subscripts 1 and 2 refer to two successive stages of eddy growth. *Feigenbaum's* constant $a$ as defined above represents the steady state emergence of fractional *Euclidean* structures. Considering dynamical eddy growth processes, *Feigenbaum's* constant $a$ also represents the steady state fractional outward mass dispersion rate and $a^2$ represents the energy flux into the environment generated by the persistent primary perturbation $W_1$. Considering both clockwise and counterclockwise rotations, the total energy flux into the environment is equal to $2a^2$. In statistical terminology, $2a^2$ represents the variance of fractal structures for both clockwise and counterclockwise rotation directions.

The probability of occurrence $P_{tot}$ of fractal domain $W_1R_1$ in the total larger eddy domain $W_nR_n$ in any (irrespective of positive or negative) direction is equal to

$$P_{tot} = \frac{W_1 R_1}{W_n R_n} = \tau^{-2n}$$

Therefore the probability $P$ of occurrence of fractal domain $W_1R_1$ in the total larger eddy domain $W_nR_n$ in any one direction (either positive or negative) is equal to

$$P = \left(\frac{W_1 R_1}{W_n R_n}\right)^2 = \tau^{-4n} \tag{9}$$

The *Feigenbaum's* constant $d$ is shown to be equal to[31,6]

$$d = \frac{W_2^4 R_2^3}{W_1^4 R_1^3} \tag{10}$$

Equation 10 represents the fractional volume intermittency of occurrence of fractal structures for each length step growth. *Feigenbaum's* constant *d* also represents the relative spin angular momentum of the growing large eddy structures as explained earlier.

Equation 1 may now be written as

$$2\frac{W^2 R^2}{w_*^2 (\mathrm{d}R)^2} = \pi \frac{W^4 R^3}{w_*^4 (\mathrm{d}R)^3} \tag{11}$$

In equation 11 d*R* equal to *r* represents the incremental growth in radius for each length step growth, i.e., *r* relates to the earlier stage of eddy growth.

The Feigenbaum's constant *d* represented by *R*/*r* is equal to

$$d = \frac{W^4 R^3}{w_*^4 r^3} \tag{12}$$

For two successive stages of eddy growth

$$d = \frac{W_2^4 R_2^3}{W_1^4 R_1^3} \tag{13}$$

From equation 1

$$W_1^2 = \frac{2}{\pi}\frac{r}{R_1} w_*^2$$
$$W_2^2 = \frac{2}{\pi}\frac{r}{R_2} w_*^2 \tag{14}$$

Therefore

$$\frac{W_2^2}{W_1^2} = \frac{R_1}{R_2} \tag{15}$$

Substituting in equation 13

$$d = \frac{W_2^4 R_2^3}{W_1^4 R_1^3} = \frac{W_2^2}{W_1^2}\frac{W_2^2 R_2^3}{W_1^2 R_1^3} = \frac{R_1}{R_2}\frac{W_2^2 R_2^3}{W_1^2 R_1^3} = \frac{W_2^2 R_2^2}{W_1^2 R_1^2} \tag{16}$$

The Feigenbaum's constant *d* represents the scale ratio $R_2/R_1$ and the inverse of the Feigenbaum's constant *d* equal to $R_1/R_2$ represents the probability $(Prob)_1$ of occurrence of length scale $R_1$ in the total fluctuation length domain $R_2$ for the first eddy growth step as given in the following

$$\left(Prob\right)_1 = \frac{R_1}{R_2} = \frac{1}{d} = \frac{W_1^2 R_1^2}{W_2^2 R_2^2} = \tau^{-4} \quad (17)$$

In general for the $n^{th}$ eddy growth step, the probability $(Prob)_n$ of occurrence of length scale $R_1$ in the total fluctuation length domain $R_n$ is given as

$$\left(Prob\right)_n = \frac{R_1}{R_n} = \frac{W_1^2 R_1^2}{W_n^2 R_n^2} = \tau^{-4n} \quad (18)$$

The above equation for probability $(Prob)_n$ also represents, for the $n^{th}$ eddy growth step, the following statistical and dynamical quantities of the growing large eddy with respect to the initial perturbation domain: (i) the statistical relative variance of fractal structures, (ii) probability of occurrence of fractal domain in either positive or negative direction, and (iii) the inverse of $(Prob)_n$ represents the organised fractal (fine scale) energy flux in the overall large scale eddy domain. Large scale energy flux therefore occurs not in bulk, but in organized internal fine scale circulation structures identified as fractals.

Substituting the *Feigenbaum's constants a* and *d* defined above (equations 8 and 10), equation 11 can be written as

$$2a^2 = \pi d \quad (19)$$

In equation 19 $\pi d$, the relative volume intermittency of occurrence contributes to the total variance $2a^2$ of fractal structures.

In terms of eddy dynamics, the above equation states that during each length step growth, the energy flux into the environment equal to $2a^2$ contributes to generate relative spin angular momentum equal to $\pi d$ of the growing fractal structures. Each length step growth is therefore associated with a factor of $2a^2$ equal to $2\tau^4$ ( ≅ 13.708203) increase in energy flux in the associated fractal domain. Ten such length step growths results in the formation of robust (self-sustaining) dominant bidirectional large eddy circulation $OR_OR_1R_2R_3R_4R_5$ (Figure 3) associated with a factor of $20a^2$ equal to 137.08203 increase in eddy energy flux. This non-dimensional constant factor characterizing successive dominant eddy energy increments is analogous to the *fine structure* constant $\propto^{-1}$ [32] observed in atomic spectra, where the spacing (energy) intervals between adjacent spectral lines is proportional to the non-dimensional *fine structure* constant equal to approximately 1/137. Further, the probability of $n^{th}$ length step eddy growth is given by $a^{-2n}$ ($\cong 6.8541^{-n}$) while the associated increase in eddy energy flux into the environment is equal to $a^{2n}$ (($\cong 6.8541^{n}$). Extreme events occur for large number of length step growths *n* with small probability of occurrence and are associated with large energy release in the fractal domain. Each length step growth is associated with one-tenth of *fine structure constant* energy increment equal to $2a^2$ ($\propto^{-1}/10 \cong 13.7082$) for bidirectional eddy circulation, or equal to one-twentieth of *fine structure constant* energy increment equal to $a^2$ ($\propto^{-1}/20 \cong 6.8541$) in any one direction, i.e., positive or negative. The energy increase between two successive eddy length step growths may be expressed as a function of $(a^2)^2$, i.e., proportional to the square of the *fine structure constant* $\propto^{-1}$. In the spectra of many atoms, what appears with coarse observations to be a single spectral line proves, with finer observation, to be a group of two or more closely spaced lines. The spacing of these fine-structure lines relative to the coarse spacing in the spectrum is proportional to the square of *fine* *structure constant*, for which reason this combination is called the *fine-structure constant*. We now know that the significance of the *fine-structure constant* goes beyond atomic spectra[32].

It was shown at equation 5 (Section 3.2.2) above that the steady state emergence of fractal structures in fluid flows is equal to $1/k$ ($=\tau^2$) and therefore the *Feigenbaum's constant a* is equal to

$$a = \tau^2 = \frac{1}{k} = 2.62 \tag{20}$$

### 3.2.4 Universal Feigenbaum's constants and power spectra of fractal fluctuations

The power spectra of fluctuations in fluid flows can now be quantified in terms of universal *Feigenbaum's constant a* as follows.

The normalized variance and therefore the statistical probability distribution is represented by (from equation 9)

$$P = a^{-2t} \tag{21}$$

In equation 21 $P$ is the probability density corresponding to normalized standard deviation $t$. The graph of $P$ versus $t$ will represent the power spectrum. The slope $S$ of the power spectrum is equal to

$$S = \frac{\mathrm{d}P}{\mathrm{d}t} \approx -P \tag{22}$$

The power spectrum therefore follows inverse power law form, the slope decreasing with increase in $t$. Increase in $t$ corresponds to large eddies (low frequencies) and is consistent with observed decrease in slope at low frequencies in dynamical systems.

The probability distribution of fractal fluctuations (equation 18) is therefore the same as variance spectrum (equation 21) of fractal fluctuations.

The steady state emergence of fractal structures for each length step growth for any one direction of rotation (either clockwise or anticlockwise) is equal to

$$\frac{a}{2} = \frac{\tau^2}{2}$$

since the corresponding value for both direction is equal to $a$ (equations. 5 and 20 ).

The emerging fractal space-time structures have moment coefficient of kurtosis given by the fourth moment equal to

$$\left(\frac{\tau^2}{2}\right)^4 = \frac{\tau^8}{16} = 2.9356 \approx 3$$

The moment coefficient of skewness for the fractal space-time structures is equal to zero for the symmetric eddy circulations. Moment coefficient of kurtosis equal to 3 and moment coefficient of skewness equal to *zero* characterize the statistical normal distribution. The model predicted power law distribution for fractal fluctuations is close to the Gaussian distribution.

### 3.2.5 Power spectrum and probability distribution of fractal fluctuations

The relationship between *Feigenbaum's constant a* and power spectra may also be derived as follows.

The steady state emergence of fractal structures is equal to the *Feigenbaum's constant a* (equations 5 and 20). The relative variance of fractal structure which also represents the probability *P* of occurrence of bidirectional fractal domain for each length step growth is then equal to $1/a^2$. The normalized variance $\frac{1}{a^{2n}}$ will now represent the statistical probability density for the $n^{th}$ step growth according to model predicted quantumlike mechanics for fluid flows. Model predicted probability density values *P* are computed as

$$P = \frac{1}{a^{2n}} = \tau^{-4n} \tag{23}$$

or

$$P = \tau^{-4t} \tag{24}$$

In equation 24 *t* is the normalized standard deviation (equation 6). The model predicted *P* values corresponding to normalised deviation *t* values less than 2 are slightly less than the corresponding statistical normal distribution values while the *P* values are noticeably larger for normalised deviation *t* values greater than 2 (Table 2 and Figure 4) and may explain the reported *fat tail* for probability distributions of various physical parameters[11]. The model predicted *P* values plotted on a linear scale (Y-axis) shows close agreement with the corresponding statistical normal probability values as seen in Figure 4 (left side). The model predicted *P* values plotted on a logarithmic scale (Y-axis) shows *fat tail* distribution for normalised deviation *t* values greater than 2 as seen in Figure 4 (right side).

Table 2 Model predicted and statistical normal probability density distributions

| growth step | normalized deviation | cumulative probability densities (%) | |
|---|---|---|---|
| *n* | *t* | model predicted $P = \tau^{-4t}$ | statistical normal distribution |
| 1 | 1 | 14.5898 | 15.8655 |
| 2 | 2 | 2.1286 | 2.2750 |
| 3 | 3 | 0.3106 | 0.1350 |
| 4 | 4 | 0.0453 | 0.0032 |
| 5 | 5 | 0.0066 | ≈ 0.0 |

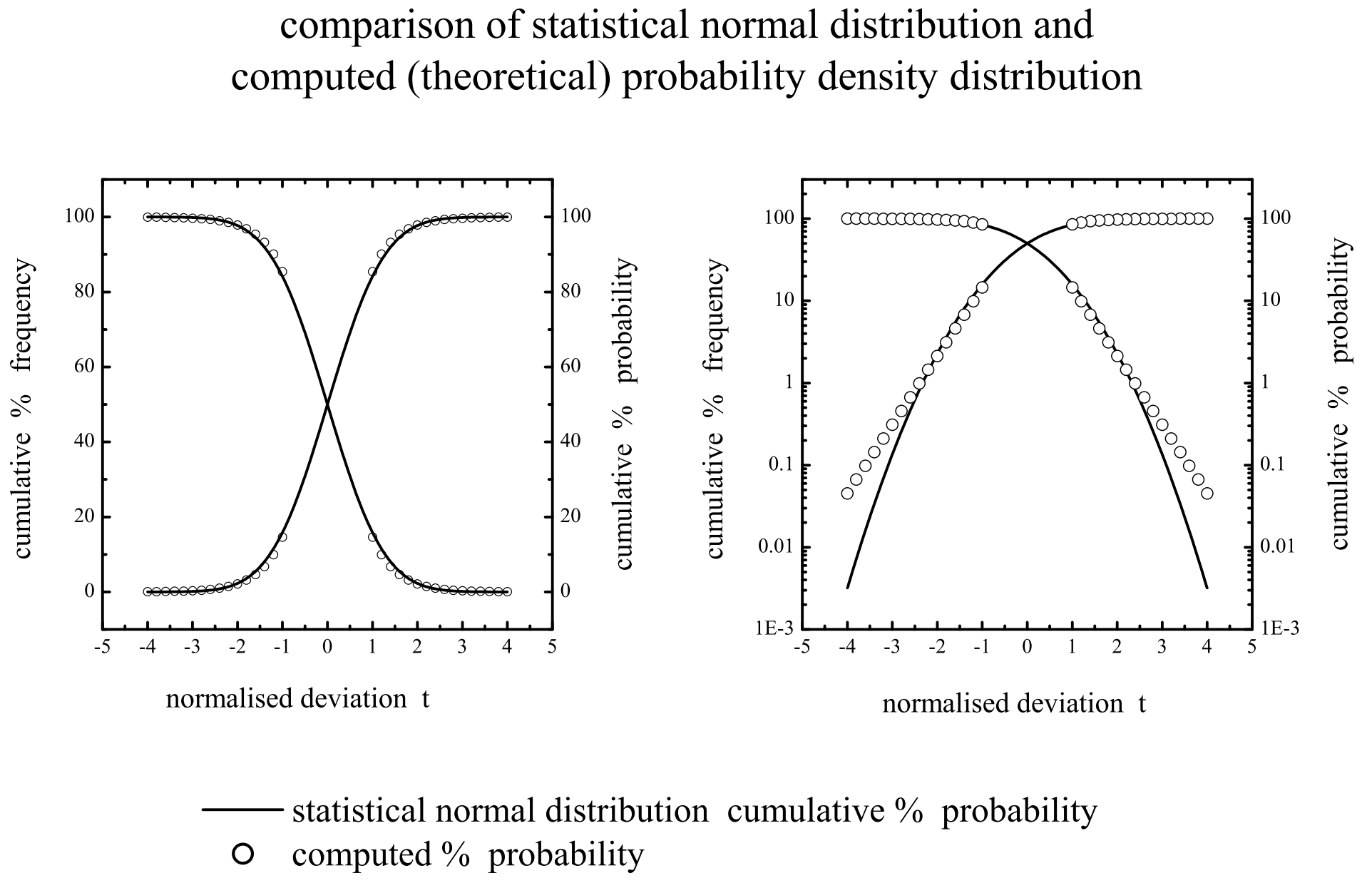

Figure 4 Comparison of statistical normal distribution and computed (theoretical) probability density distribution. The same figure is plotted on the right side with logarithmic scale for the probability axis (Y-axis) to show clearly that for normalized deviation *t* values greater than 2 the computed probability densities are greater than the corresponding statistical normal distribution values.

# 4 General Systems Theory and Classical Statistical Physics

Kinetic theory of ideal gases is a study of systems consisting of a great number of molecules, which are considered as bodies having a small size and mass[33]. Classical statistical methods of investigation[33-40] are employed to estimate average values of quantities characterizing aggregate molecular motion such as mean velocity, mean energy etc., which determine the macro-scale characteristics of gases. The mean properties of ideal gases are calculated with the following assumptions. (1) The intra-molecular forces are completely absent instead of being small. (2) The dimensions of molecules are ignored, and considered as material points. (3) The above assumptions imply the molecules are completely free, move rectilinearly and uniformly as if no forces act on them. (4) The ceaseless chaotic movements of individual molecules obey Newton's laws of motion.

For any system large or small in thermal equilibrium at temperature $T$, the probability $P$ of being in a particular state at energy $E$ is proportional to $e^{-\frac{E}{k_B T}}$ where $k_B$ is the *Boltzmann's constant*. This is called the *Boltzmann distribution* for molecular energies and may be written as

$$P \propto e^{-\frac{E}{k_B T}} \tag{25}$$

The physical concepts of the general systems theory (Section 3) enables to derive[17] *Boltzmann distribution* as shown in the following.

The r.m.s circulation speed *W* of the large eddy follows a logarithmic relationship with respect to the length scale ratio *z* equal to *R/r* (equation 4 ) as given below

$$W = \frac{w_*}{k} \log z$$

In the above equation the variable *k* represents for each step of eddy growth, the fractional volume dilution of large eddy by turbulent eddy fluctuations carried on the large eddy envelope[4] and is given as

$$k = \frac{w_* r}{WR}$$

Substituting for *k* in equation (4) we have

$$W = w_* \frac{WR}{w_* r} \log z = \frac{WR}{r} \log z$$

*and*

$$\frac{r}{R} = \log z$$

Therefore

$$z = \frac{R}{r} = e^{\frac{r}{R}} \quad or \quad \frac{r}{R} = e^{-\frac{r}{R}} \qquad (26)$$

The ratio *r/R* represents the fractional probability *P* of occurrence of small-scale fluctuations (*r*) in the large eddy (*R*) environment. Considering two large eddies of radii $R_1$ and $R_2$ ($R_2$ greater than $R_1$) and corresponding r.m.s circulation speeds $W_1$ and $W_2$ which grow from the same primary small-scale eddy of radius *r* and r.m.s circulation speed $w_*$ we have from equation (1)

$$\frac{R_1}{R_2} = \frac{W_2^2}{W_1^2}$$

From equation (26)

$$\frac{R_1}{R_2} = e^{-\frac{R_1}{R_2}} = e^{-\frac{W_2^2}{W_1^2}} \qquad (27)$$

The square of r.m.s circulation speed $W^2$ represents eddy kinetic energy. Following classical physical concepts[33] the primary (small-scale) eddy energy may be written in terms of the eddy environment temperature *T* and the *Boltzmann's constant* $k_B$ as

$$W_1^2 \propto k_B T$$

Representing the larger scale eddy energy as $E$

$$W_2^2 \propto E$$

The length scale ratio $R_1/R_2$ therefore represents fractional probability $P$ of occurrence of large eddy energy $E$ in the environment of the primary small-scale eddy energy $k_B T$ (equation 18). The expression for $P$ is obtained from equation (27) as

$$P \propto e^{-\frac{E}{k_B T}}$$

The above is the same as the *Boltzmann's equation* (equation 25).

The derivation of *Boltzmann's equation* from general systems theory concepts visualises the eddy energy distribution as follows: (1) The primary small-scale eddy represents the molecules whose eddy kinetic energy is equal to $k_B T$ as in classical physics. (2) The energy pumping from the primary small-scale eddy generates growth of progressive larger eddies[4]. The r.m.s circulation speeds $W$ of larger eddies are smaller than that of the primary small-scale eddy (equation 2). (3) The space-time *fractal* fluctuations of molecules (atoms) in an ideal gas may be visualised to result from an eddy continuum with the eddy energy $E$ per unit volume relative to primary molecular kinetic energy ($k_B T$) decreasing progressively with increase in eddy size.

The eddy energy probability distribution ($P$) of fractal space-time fluctuations also represents the *Boltzmann distribution* for each stage of hierarchical eddy growth and is given by equation (24) derived earlier, namely

$$P = \tau^{-4t}$$

The general systems theory concepts are applicable to all space-time scales ranging from microscopic scale quantum systems to macroscale real world systems such as atmospheric flows.

# 5 Data and Analysis

## 5.1 Data

Gridded temperature anomalies for mean temperatures obtained from the GHCN V2 (http://www.ncdc.noaa.gov/oa/climate/ghcn-monthly/index.php and ftp://ftp.ncdc.noaa.gov/pub/data/ghcn/v2/) monthly temperature data sets for 129 years (1880 to 2008) the months January to May and for 128 years (1880 to 2008) for the months June to December were used for the study. Details of the data set as given in the 'README_GRID_TEMP' text document are as follows. GHCN homogeneity adjusted data was the primary source for developing the gridded fields. In grid boxes without homogeneity adjusted data, GHCN raw data was used to provide additional coverage when possible. Each month of data consists of 2592 gridded data points produced on a 5 X 5 degree basis for the entire globe (72 longitude X 36 latitude grid boxes).

Gridded data for every month from January 1880 to the most recent month (May 2008 in the present study) is available. The data are temperature anomalies in degrees

Celsius. Each gridded value was multiplied by 100 and written to file as an integer. Missing values are represented by the value -32768.

There are two options: grid_1880_YYYY.dat.gz (used in the present study), where YYYY is the current year and the values are calculated using the first difference method, and anom-grid-1880-current.dat.gz which uses the "anomaly" method.

The data are formatted by year, month, latitude and longitude. There are twelve longitude grid values per line, so there are 6 lines (72/12 = 6) for each of the 36 latitude bands. Longitude values are written from 180 W to 180 E, and latitude values from 90 N to 90 S. Data for each month is preceded by a label containing the month and year of the gridded data.

```
for year = begyr to endyr
 for month = 1 to 12
   format(2i5) month,year
   for ylat = 1 to 36 (85-90N,80-85N,...,80-85S,85-90S)
     format(12i5) 180-175W,175-170W,...,130-125W,125-120W
     format(12i5) 120-115W,175-170W,...,70-65W,65-60W
     format(12i5) 60-55W,55-50W,...,10-5W,5-0W
     format(12i5) 0-5E,5-10E,...,50-55E,55-60E
     format(12i5) 60-65E,65-70E,...,110-115E,115-120E
     format(12i5) 120-125E,125-130E,...,170-175E,175-180E
```

Each file has been compressed using 'gzip'. They can be uncompressed with 'WinZip' for those using Windows 95 (and above) or with 'gzip' from most UNIX platforms. The FORTRAN utility program 'read_gridded.f' can be downloaded to assist in extracting data of interest. This program allows the user to extract non-missing values for selected months and write the data to an ascii output file. The latitude and longitude of the center of each corresponding grid box accompanies each gridded value in the output file.

These anomalies were calculated with respect to the period 1961 - 1990 using the First Difference Method, an approach developed to maximize the use of available station records[41-43]. The First Difference Method involves calculating a series of calendar-month differences in temperature between successive years of station data (FDyr = Tyr - Tyr-1). For example, when creating a station's first difference series for mean February temperature, we subtract the station's February 1880 temperature from the station's February 1881 temperature to create a February 1881 first difference value. First difference values for subsequent years are calculated in the same fashion by subtracting the station's preceding year temperature for all available years of station data.

For each year and month the sum of the 'first difference' value of all stations located within the appropriate 5 X 5 degree box was determined and divided by the total number of stations in the box to get an unweighted first difference value for each grid box. Next calculate a cumulative sum of these gridded first difference values for all years from 1880 to 1998 to produce a time series for each grid box. The cumulative sum is calculated for each grid box and each month of gridded first difference data independently through time. Each grid box time series is then adjusted to create anomalies with respect to the period 1961 - 1990.

This gridded data set was developed to produce the most accurate time series possible. However, this required to treat months and grid boxes independently through time. The use of this data is most appropriate for analyzing the change in temperature within a particular grid box, or set of grid boxes, over a span of years. If one is more interested in analyzing temperature changes within individual years, e.g., the change in temperature between February and March, 1908, or between two regions in 1908, it is recommended that the GHCN station data be used directly.

## 5.2 Analyses and Results

### 5.2.1 Frequency distribution

Each data set (month-wise temperature time series for the period 1800 to 2008 for the months January to May and for the period 1800 to 2007 for the months June to December) was represented as the frequency of occurrence $f(i)$ in a suitable number $n$ of class intervals $x(i)$, $i=1, n$ covering the range of values from *minimum* to the *maximum* in the data set. The class interval $x(i)$ represents dataset values in the range $x(i) \pm \Delta x$, where $\Delta x$ is a constant. The average *av* and standard deviation *sd* for the data set is computed as

$$av = \frac{\sum_{1}^{n}\left[x(i) \times f(i)\right]}{\sum_{1}^{n} f(i)}$$

$$sd = \frac{\sum_{1}^{n}\left\{\left[x(i) - av\right]^2 \times f(i)\right\}}{\sum_{1}^{n} f(i)}$$

The *normalized deviation t* values for class intervals $t(i)$ were then computed as

$$t(i) = \frac{x(i) - av}{sd}$$

The cumulative percentage probabilities of occurrence $cmax(i)$ and $cmin(i)$ corresponding to the *normalized deviation t* values were then computed starting respectively from the maximum ($i=n$) and minimum ($i=1$) class interval values as follows.

$$cmax(i) = \frac{\sum_{n}^{i}\left[x(i) \times f(i)\right]}{\sum_{1}^{n}\left[x(i) \times f(i)\right]} \times 100.0$$

$$cmin(i) = \frac{\sum_{1}^{i}\left[x(i) \times f(i)\right]}{\sum_{1}^{n}\left[x(i) \times f(i)\right]} \times 100.0$$

The *cmax* and *cmin* distributions with respect to the *normalized deviation t* values were computed for each month for all grid points including grid points which did not have continuous time series data. The total number of grid points available for the study for the months January to December is given in Figure 5. The average and standard deviation of *cmax(i)* and *cmin(i)* for each *normalized deviation t(i)* values were then computed for each month from the corresponding distributions for all the available grid points for the month. The average cumulative percentage probability values *cmax*($i$) and *cmin*($i$) plotted with respect to corresponding *normalized deviation* $t(i)$ values with linear axes are shown in Figure 6 and on logarithmic scale for the probability axis in the tail region, i.e. *normalized deviation* $t(i)$ values greater than 2 in Figure 7 for positive extremes $t(i)$ = 2 to 4, and in Figure 8 for negative extremes $t(i)$ = -2 to -4. The standard deviation of each mean *cmax*($i$) and *cmin*($i$) value is shown as a vertical error bar on either side of the mean in Figures 6 to 8. The figures also contain the statistical normal distribution and the computed theoretical distribution (equation 24) for comparison. The figures show clearly the appreciable positive departure of observed probability densities from the statistical normal distribution for extreme values at normalised deviation $t$ values more than 2 (Figures. 7 and 8). The observed extreme values corresponding to $t(i)$ values greater than 2 for *cmax*($i$) and *cmin*($i$) distributions were compared for 'goodness of fit' with computed theoretical distribution and statistical normal distribution as follows. For *cmax*($i$) and *cmin*($i$) values, where standard deviation is available (no of observed values more than one), if the observed distribution included the computed theoretical (statistical normal) distribution within twice the standard deviation on either side of the mean then it was assumed to be the same as the computed theoretical (statistical normal) distribution at 5% level of significance within measurement errors. The number of observed distribution values which included the computed theoretical values and/or the statistical normal distribution values within twice the standard deviation on either side of the mean was determined. The total and percentage numbers of observed extreme values same as computed theoretical and statistical normal distributions are given in Figure 9 for positive and negative tail regions (normalized deviation $t$ greater than 2) for the 12 months January to December. The percentage number of observed extreme value points same as model predicted computed is more than the percentage number of extreme value points same as statistical normal distribution for all the 12 months. The observed distribution is closer to the model predicted theoretical than the statistical normal distribution.

### 5.2.2 Continuous periodogram power spectral analyses

The power spectra of frequency distribution of monthly mean data sets were computed accurately by an elementary, but very powerful method of analysis developed by Jenkinson[44] which provides a quasi-continuous form of the classical periodogram allowing systematic allocation of the total variance and degrees of freedom of the data series to logarithmically spaced elements of the frequency range (0.5, 0). The cumulative percentage contribution to total variance was computed starting from the high frequency side of the spectrum. The power spectra were plotted as cumulative percentage contribution to total variance versus the normalized standard deviation $t$ equal to $\left(\log L/\log T_{50}\right)-1$ where $L$ is the period in years and $T_{50}$ is the period up to which the cumulative percentage contribution to total variance is equal to 50 (equation 7). The statistical *Chi-Square* test[45] was applied to determine the 'goodness of fit' of variance spectra with statistical normal distribution which is close to model predicted variance spectrum (equations 21 and 24). The average and standard deviation of cumulative percentage contribution to total variance for each *normalized deviation* $t(i)$ values were then computed for each month from the corresponding distributions for all the available grid points for the month. The average power spectra with corresponding standard deviations for each of the 12 months are plotted in Figure 10. The total number of grid points and the percentage number of grid points with variance spectra same as statistical normal distribution are shown in Figure 11. The mean power spectra follow closely the statistical normal distribution for the twelve months (Figure 10). The power spectra mostly cover the region for *normalized deviation t* less than 2 where the model predicted theoretical distribution is close to the statistical normal distribution. A majority (more than 90%) of the power spectra follow closely statistical normal distribution (Figure 11) consistent with model prediction of quantumlike chaos, i.e., variance or square of eddy amplitude represents the probability distribution, a signature of quantum systems.

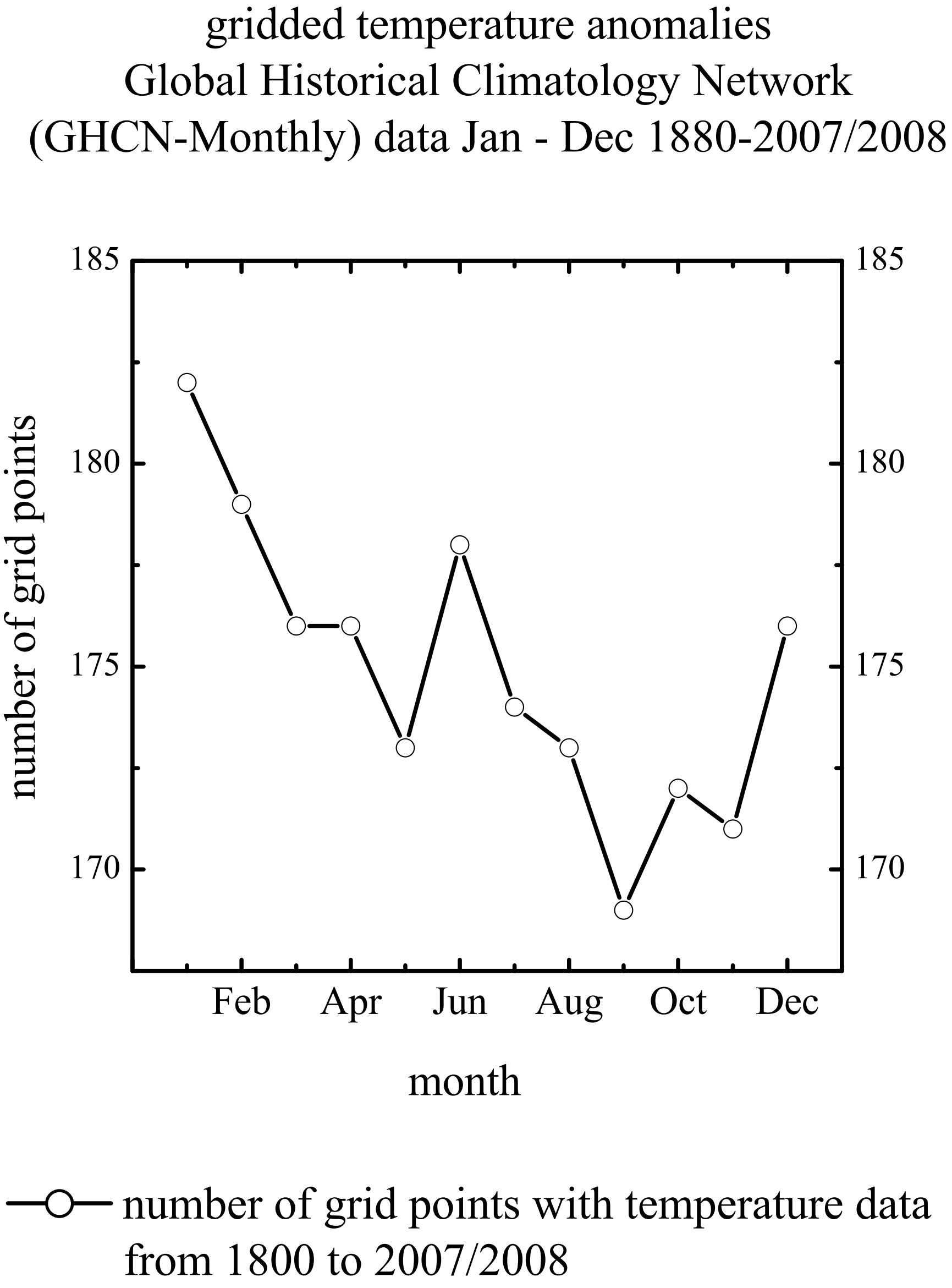

Figure 5 The total number of grid points available for the study for the months January to December.

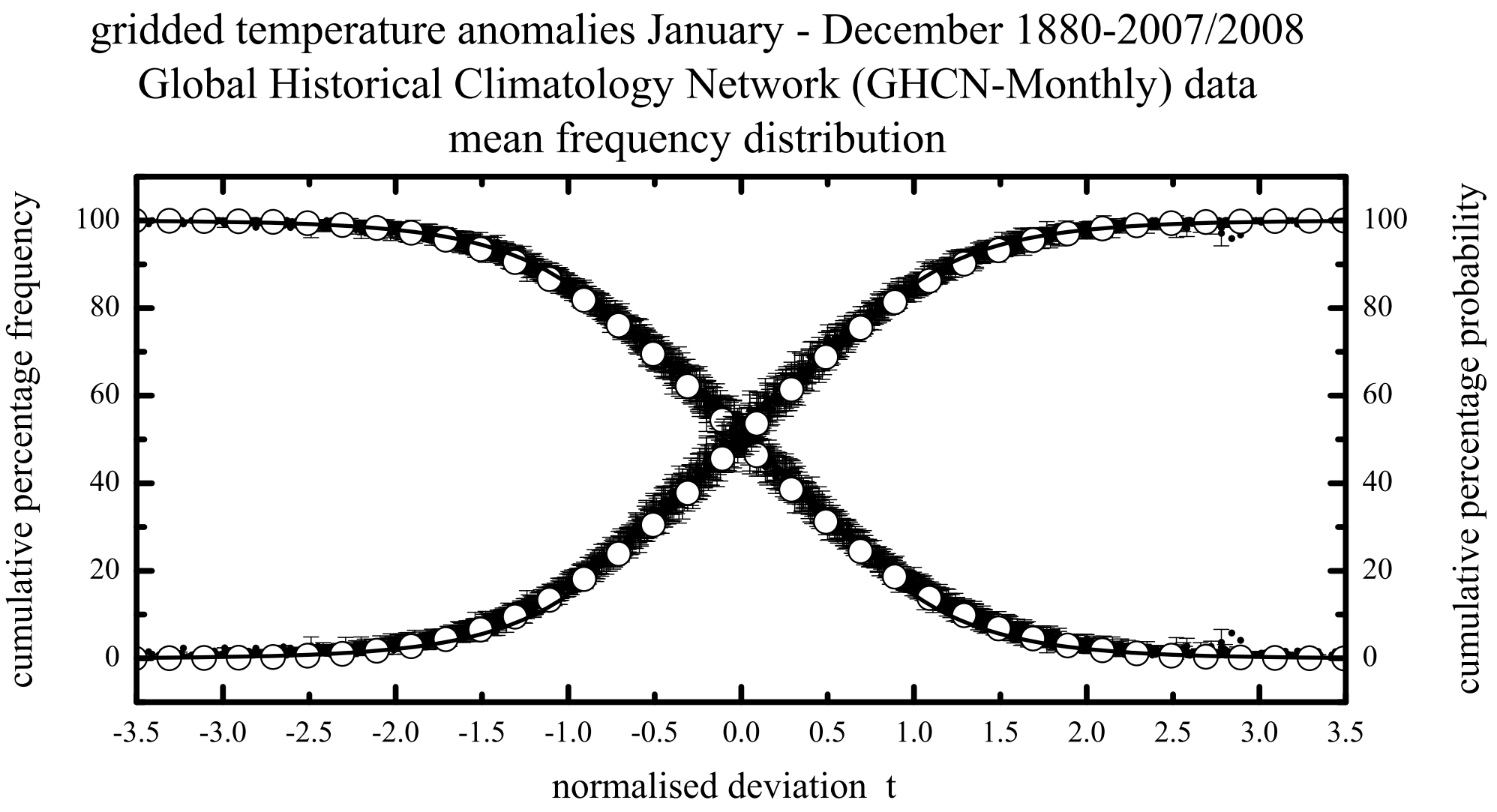

• gridded temperature anomalies Jan - Dec 1880-2007/2008 mean frequency distribution error bars (vertical) indicate one standard deviation

○ statistical normal distribution —— computed (theoretical) distribution

Figure 6 Average cumulative percentage probability distribution.

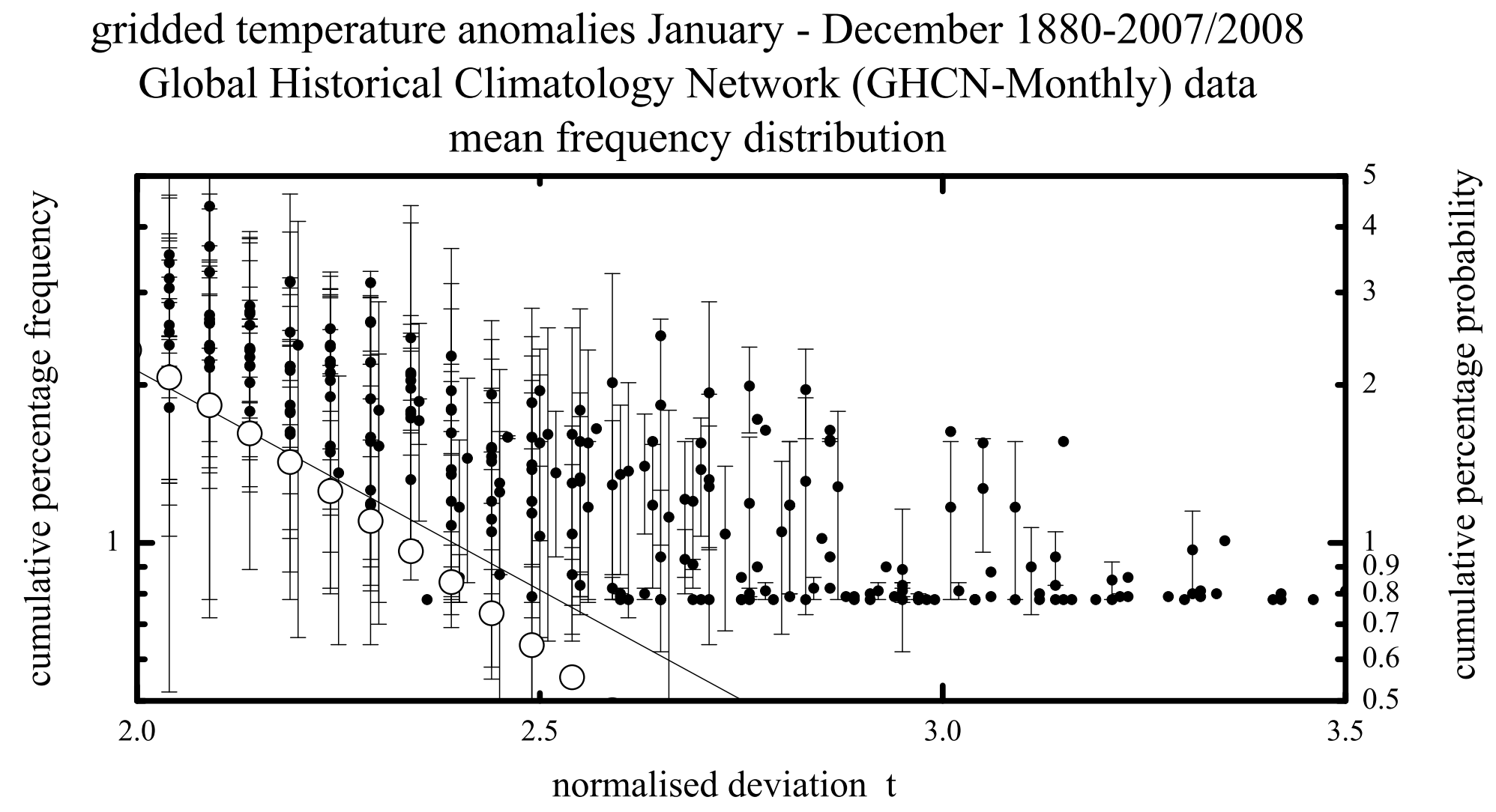

• gridded temperature anomalies Jan - Dec 1880-2007/2008 mean frequency distribution error bars (vertical) indicate one standard deviation

○ statistical normal distribution —— computed (theoretical) distribution

Figure 7 Average cumulative percentage probability values on logarithmic scale for the probability axis in the positive tail region (extreme events), i.e., *normalized deviation t* values greater than 2.

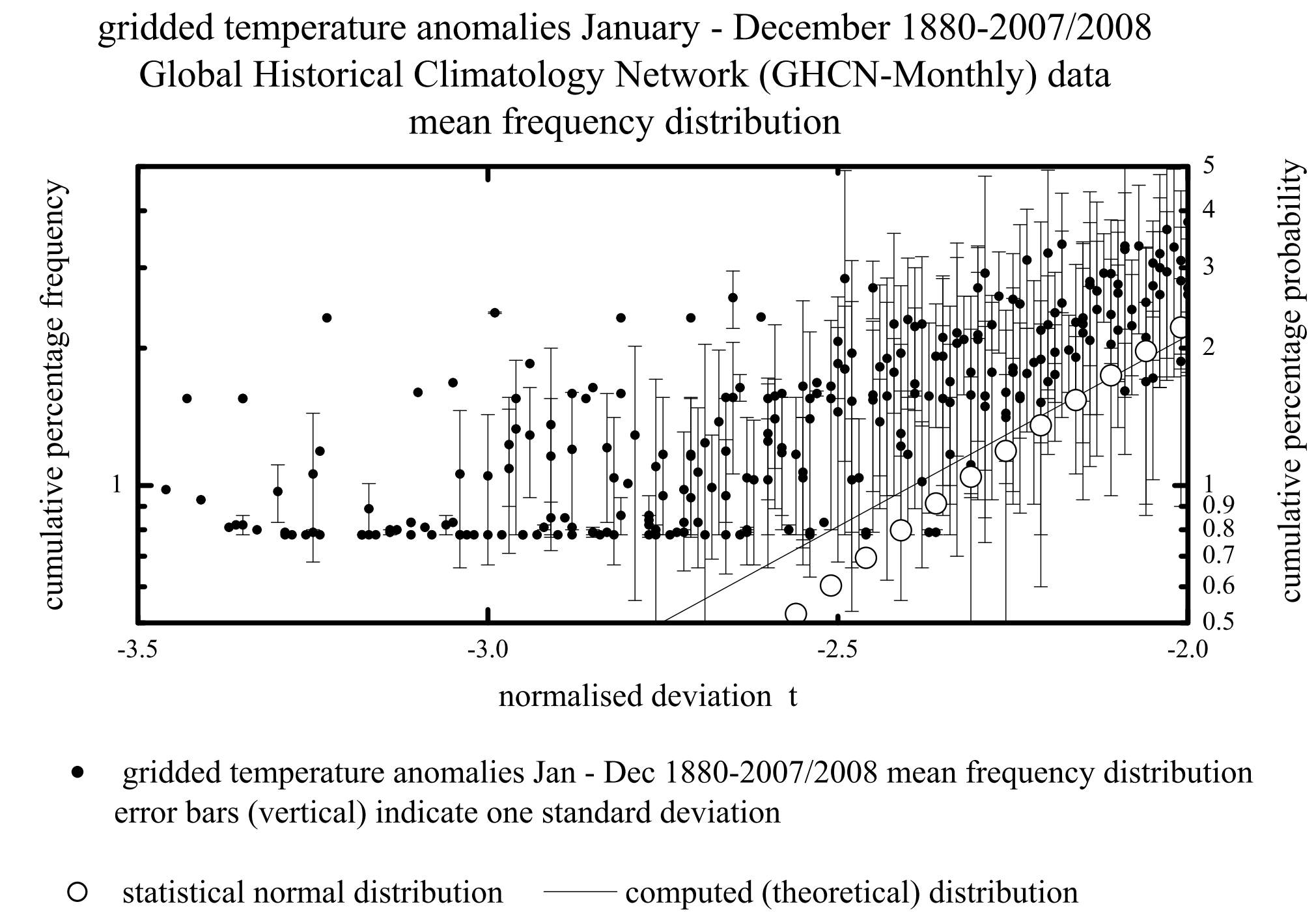

Figure 8 Average cumulative percentage probability values on logarithmic scale for the probability axis in the negative tail region (extreme events), i.e. *normalized deviation t* values greater than -2.

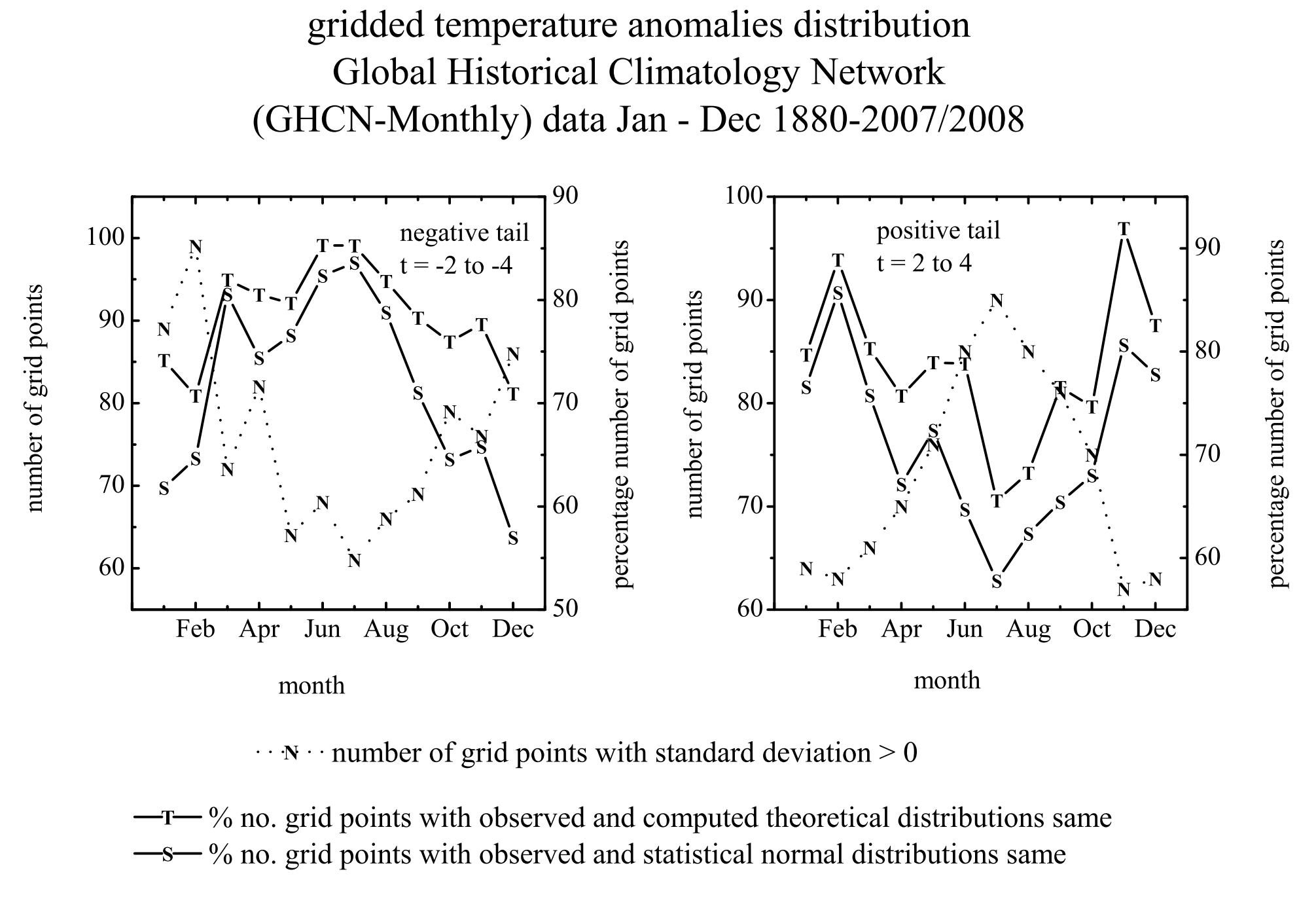

Figure 9 The total and percentage numbers of observed extreme values same as model predicted theoretical and statistical normal distributions for positive and negative tail regions (normalized deviation *t* greater than 2) for the 12 months January to December.

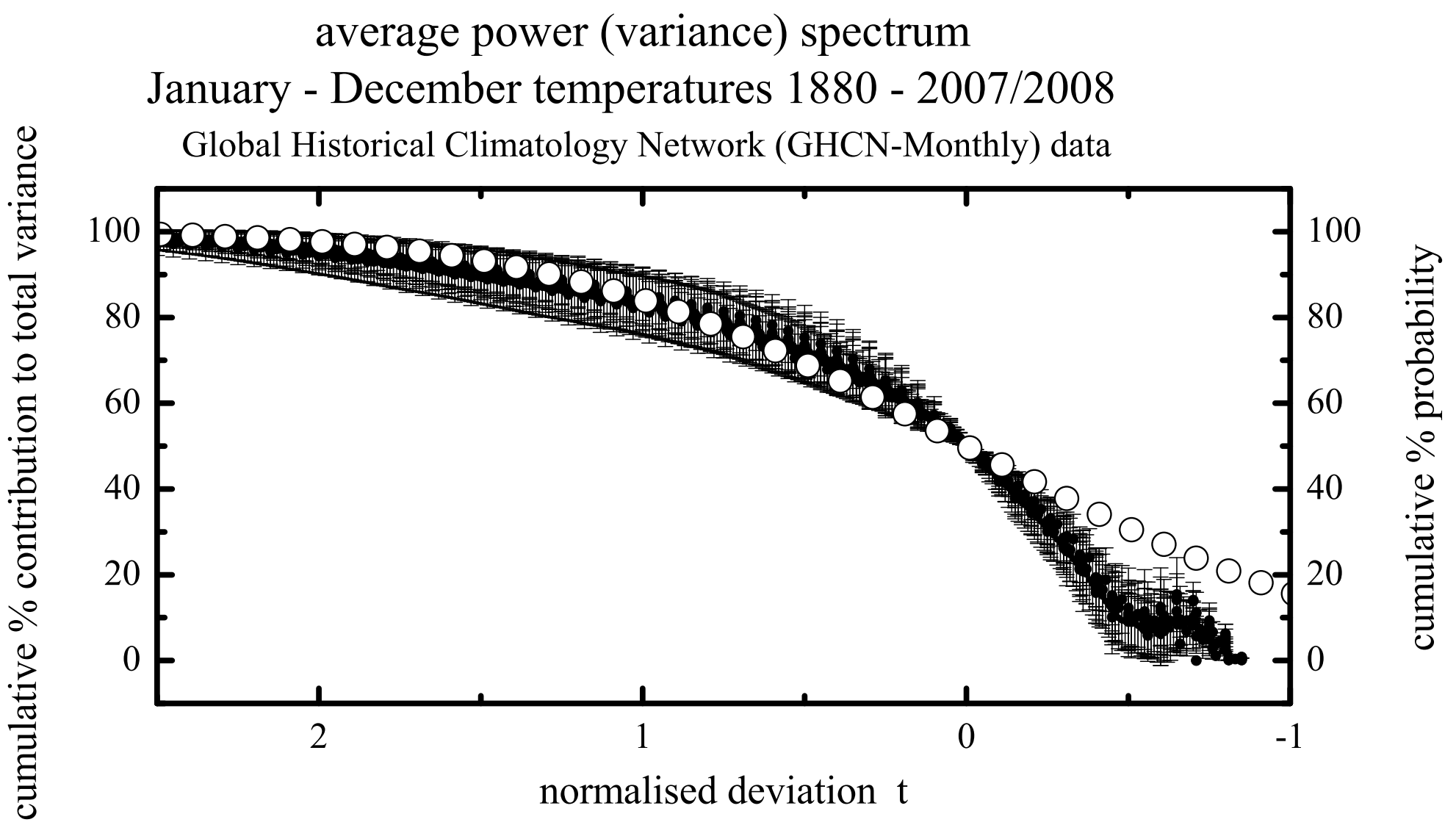

- average power (variance) spectrum January - December 1880 - 2007/2008
  error bars indicate one standard deviation (+/-)

○ statistical normal probability distribution

Figure 10 Average power (variance) spectrum with vertical error bars indicating one standard deviation on either side of the mean. The statistical normal distribution is also shown in the figure for comparison.

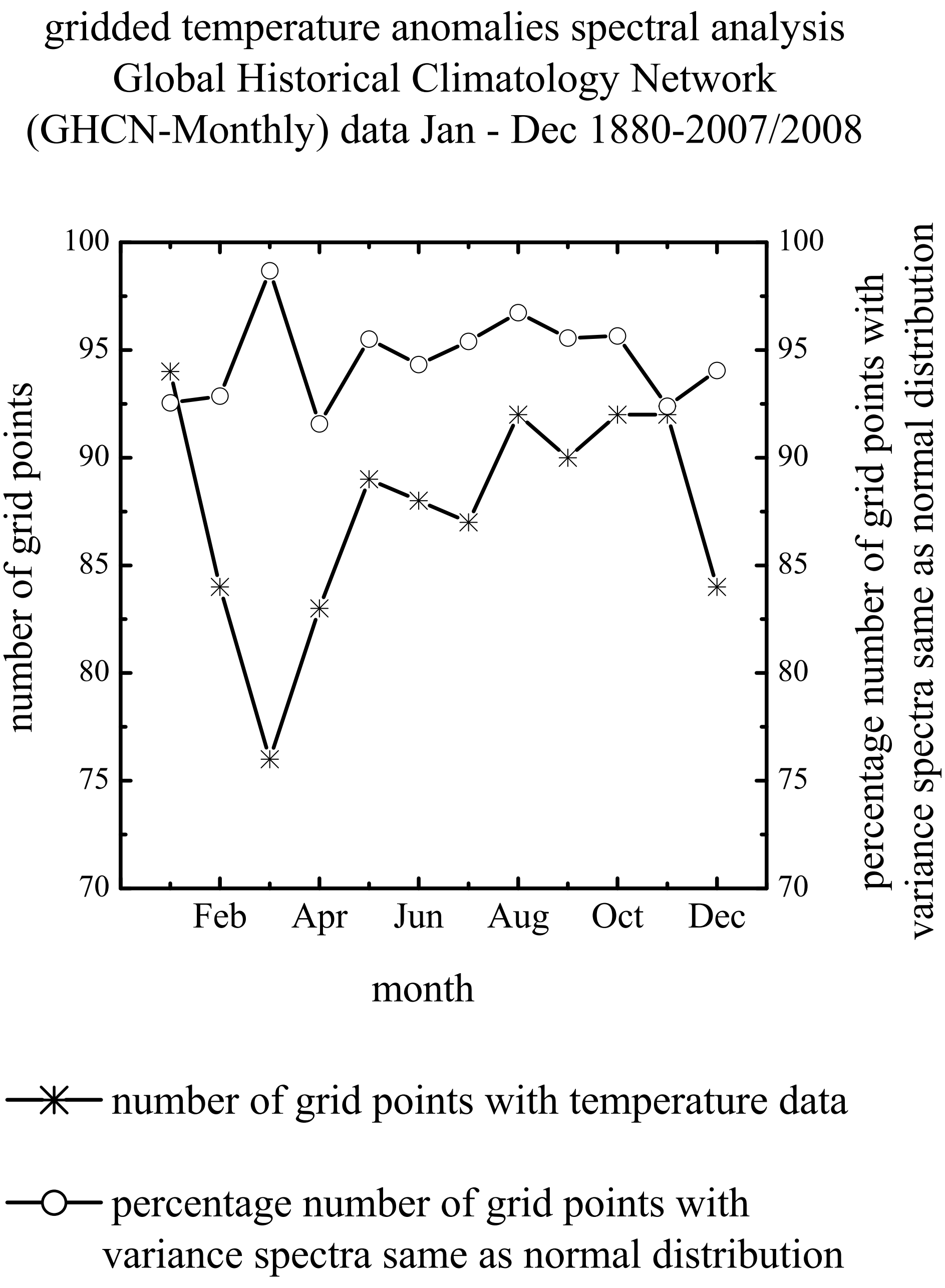

Figure 11 The total number of grid points and the percentage number of grid points with variance spectra same as statistical normal distribution.

## 6 Discussion and Conclusions

Dynamical systems in nature exhibit selfsimilar fractal fluctuations for all space-time scales and the corresponding power spectra follow inverse power law form signifying long-range space-time correlations identified as self-organized criticality.

The physics of self-organized criticality is not yet identified. The Gaussian probability distribution used widely for analysis and description of large data sets is found to significantly underestimate the probabilities of occurrence of extreme events such as stock market crashes, earthquakes, heavy rainfall, etc. Further, the assumptions underlying the normal distribution such as fixed mean and standard deviation, independence of data, are not valid for real world fractal data sets exhibiting a scale-free power law distribution with fat tails. It is important to identify and quantify the fractal distribution characteristics of dynamical systems for predictability studies.

A recently developed general systems theory for fractal space-time fluctuations[4-7] shows that the larger scale fluctuation can be visualized to emerge from the space-time averaging of enclosed small scale fluctuations, thereby generating a hierarchy of selfsimilar fluctuations manifested as the observed eddy continuum in power spectral analyses of fractal fluctuations.

The model predictions are as follows.

- The probability distribution function $P$ of fractal fluctuations follow inverse power law form $\tau^{-4t}$ where $\tau$ is the *golden mean*, and $t$, the normalized deviation is equal to ($x$-$av$/$sd$) where $av$ and $sd$ are respectively the average and standard deviation of the distribution. The predicted distribution is close to the Gaussian distribution for small-scale fluctuations (normalized deviation $t$ less than 2), but exhibits *fat long tail* for large-scale fluctuations (normalized deviation $t$ more than 2) with higher probability of occurrence than predicted by Gaussian distribution. There is always a non-zero probability of occurrence of very large amplitude, damage causing extreme events.
- The inverse of the probability distribution function, i.e., $1/P$ equal to $\tau^{4t}$ represents the domain size extent of the internal fine scale (fractal) fluctuations of amplitude $t$ (normalized deviation). High intensity extreme events corresponding to $t$ values more than 2 occur with less probability over a larger size domain and are associated with widespread damage and loss such as in heavy rainfall, earthquakes, traffic jams, etc.
- The power spectra of fractal fluctuations (Section 3) also follow inverse power law form $\tau^{-4t}$ and $t$, the normalized deviation is expressed in terms of component periodicities $L$ as $t = \{(\log L/\log T_{50})-1\}$ where $T_{50}$ is the period upto which the cumulative percentage contribution to total variance is equal to 50. Since the power (variance, i.e., square of eddy amplitude) spectrum also represents the probability densities as in the case of quantum systems such as the electron or photon, fractal fluctuations exhibit quantumlike chaos.
- The precise geometry of the quasiperiodic Penrose tiling pattern underlie fractal fluctuations tracing out robust (self-sustaining) dominant bidirectional large eddy circulation $OR_OR_1R_2R_3R_4R_5$ (Figure 3) associated with a factor of $20a^2$ ($20\tau^4$) equal to 137.08203 increase in eddy energy flux. This non-dimensional dominant eddy energy flux is analogous to and almost equal to the *fine structure constant* $\propto^{-1}$ of atomic spectra; the energy spacing intervals between successive atomic spectral lines. Further, the energy increase between two successive fine strucure eddy length step growths internal to the dominant large eddy domain may be expressed as a function of $(a^2)^2$, i.e., proportional to the square of the *fine structure constant* $\propto^{-1}$. In the spectra of many atoms, what appears with coarse observations to be a single spectral line proves, with finer observation, to be a group of two or more closely spaced lines. The spacing of these fine-structure lines relative to the coarse spacing in the spectrum is proportional to the square of *fine structure constant*, for which reason this combination is called the *fine-structure constant*[32].

Analysis of historic (1880 -2008) data sets of global monthly mean temperature (GHCN V2) time series shows that the data follow closely, but not exactly the statistical normal distribution in the region of normalized deviations *t* less than 2 (Figure 6). For normalized deviations *t* greater than 2, the data exhibit significantly larger probabilities as compared to the normal distribution and closer to the model predicted probability distribution (Figures 7 and 8). A simple *t* test for 'goodness of fit' of the extreme values (normalized deviation t > 2) of the observed distribution with model predicted (theoretical) and also the statistical normal distribution shows that more number of data points exhibit significant (at 5% level) 'goodness of fit' with the model predicted (theoretical) distribution than with the normal distribution (Figure 9). The mean power spectra follow closely the statistical normal distribution for the twelve months (Figure 10). The power spectra mostly cover the region for *normalized deviation t* less than 2 where the model predicted theoretical distribution is close to the statistical normal distribution. A majority (more than 90%) of the power spectra follow closely statistical normal distribution (Figure 11) consistent with model prediction of quantumlike chaos, i.e., variance or square of eddy amplitude represents the probability distribution, a signature of quantum systems. The model predicted and observed universal spectrum for interannual variability rules out linear secular trends in global monthly mean temperatures. Global warming results in intensification of fluctuations of all scales and manifested immediately in high frequency fluctuations. The general systems theory, originally developed for turbulent fluid flows, provides universal quantification of physics underlying fractal fluctuations and is applicable to all dynamical systems in nature independent of its physical, chemical, electrical, or any other intrinsic characteristic.

## Acknowledgement

The author is grateful to Dr. A. S. R. Murty for encouragement during the course of this study.